# WindCline: Sloping Wind Tunnel for Characterizing Flame Behavior Under Variable Inclines and Wind Conditions

Amanda S. Makowiecki[1,†], Sean C. Coburn[1,†,*], Samantha Sheppard[2], Brendan Bitterlin[1], Timothy Breda[1], Abdul Dawlatzai[1], Robert Giannella[1], Alexandra Jaros[1], Christopher Kling[1], Eric Kolb[1], Caelan Lapointe[1], Sam Simons-Wellin[1], Hope A. Michelsen[1], John W. Daily[1], Michael Hannigan[1], Peter E. Hamlington[1], John Farnsworth[2], and Gregory B. Rieker[1,*]

[1]Paul M. Rady Department of Mechanical Engineering, University of Colorado Boulder, Boulder, CO
[2]Ann and H.J. Smead Department of Aerospace Engineering Sciences, University of Colorado Boulder, Boulder, CO
*Correspondence: coburns@colorado.edu; or greg.rieker@colorado.edu
†These authors contributed equally to this work

**Abstract**
Developing accurate computational models of wildfire dynamics is increasingly important due to the substantial and expanding negative impacts of wildfire events on human health, infrastructure, and the environment. Wildfire spread and emissions depend on a number of factors, including fuel type, environmental conditions (moisture, wind speed, etc.), and terrain/location. However, there currently exist only a few experimental facilities that enable testing of the interplay of these factors at length scales <1 m with carefully controlled and characterized boundary conditions and advanced diagnostics. Experiments performed at such facilities are required for informing and validating computational models. Here we present the design and characterization of a novel tilting wind tunnel (the 'WindCline') for studying wildfire dynamics. The WindCline is unique in that the entire tunnel platform is constructed to pivot around a central axis, which enables sloping of the entire system without compromising the quality of the flow properties. In addition, this facility has a configurable design for the test section and diffuser to accommodate a suite of advanced diagnostics to aid in the characterization of 1) the parameters needed to establish boundary conditions and 2) flame properties and dynamics. The WindCline thus allows for measurement and control of several critical wildfire variables and boundary conditions, especially at the small length scales important to the development of high-fidelity computational simulations (10 – 100 cm). Computational modeling frameworks developed and validated under these controlled conditions can expand understanding of fundamental combustion processes, promoting greater confidence when leveraging these processes in complex combustion environments.

**Introduction:**
Understanding wildfire spread is of increasing importance because of the expanding wildland urban interface and the impacts of climate change [1–6]. Incorporating environmental characteristics and conditions into wildfire computational models is critical for accurately predicting wildfire spread, identifying at-risk areas, and informing fire-suppression decisions. Ultimately, robust computational models will lead to better protection and alert systems for citizens and help to minimize impacts on human health, infrastructure, and the environment. However, the development of such models is challenging due to the complex nature of fire spread, which is dependent on variables that range in scope and scale from the molecular level (e.g., the chemical composition of the fuel) to the regional level (e.g., regional climate and terrain characteristics). Many studies on fire dynamics and spread focus on understanding the relationships between fire parameters and environmental variables through controlled experimental frameworks [7–15].

Among the environmental variables, the impact of weather and topography are of particular importance. For example, in mountainous landscapes, unexpected and rapid fire spread is possible due to the presence of high fuel loads combined with complex wind dynamics created from varied and unique terrain and strong thermal gradients [16,17]. While it is possible to incorporate these effects into computational models, it is critical to test the models under extremely well-characterized and, where possible, controlled conditions which allows the components of the computational models that handle various physics (chemical kinetics, fluid dynamics, heat transfer, etc.) to be individually tested, refined, and validated. Having a well-informed and validated basis for the models serves to improve the robustness of simulations of more complex environments which are harder to recreate in controlled experiments under known conditions.

There are numerous experimental platforms that are designed for probing various aspects of flame dynamics and fire spread for wildfire studies. Crossflow interacting with an open flame testbed is a commonly used configuration to control the inflow conditions to wildfire experiments [12,18–22]. This experimental architecture enables wildfire studies at laboratory scales (i.e., multiple meters), which begin to approach scales relevant for outdoor wildfires while maintaining the fidelity of controlled experiments (i.e., some control over boundary conditions to inform models). Another experimental configuration that targets the impacts of topography on flame dynamics and fire spread incorporates a flame testbed on a platform with an adjustable angle. Studies utilizing this type of apparatus have been performed with different configurations, including: the testbed being placed on a tilting platform in an open room [23–27], the testbed being placed at an angle within a wind tunnel [28–30], and where testbed and wind tunnel tilt together [31–34]. Depending on the study focus there exist examples both with and without the addition of a forced crossflow around the testbed. Many experiments presented in the literature focus on the effects of either forced flow or inclination on flame behavior and few investigations study these parameters together. Validating computational fluid dynamics models and creating benchmark data for their development requires accurate knowledge of the boundary conditions and experimental variables, and benefits from the ability to independently control these conditions and variables.

Here we report on the development of a novel tilting wind tunnel (WindCline) designed to accommodate flaming combustion and is configurable to enable an array of state-of-the-art diagnostics. We follow the design and characterization fundamentals matured over decades for wind-tunnel design by the aerodynamics community to create well-characterized and controlled boundary conditions. The platform will enable the development of robust computational fluid dynamics (CFD) models for small length scales (10 – 100 cm) under a range of environmental conditions.

The WindCline consists of a blow-down wind tunnel configuration (adjustable blower, flow conditioning, contraction, test section, and diffuser) assembled onto a tilting platform, which allows the entire wind tunnel to be angled while preserving the flow conditions during angled experiments. The test section is configured with a removable baseplate to enable either stationary-flame (e.g., a slot-burner) or solid-fuel experiments. Having the ability to control these different parameters allows us to isolate and characterize variables critical for understanding fire spread, including wind speed, fuel energy (i.e., fuel flow rate), and terrain angle. The facility test section is configurable to enable various measurement techniques aimed at understanding different flow and chemical physics during experiments. In this paper we present the wind tunnel design and characterization, followed by initial measurements on stationary flames and solid fuels using a suite of diagnostic approaches.

**Design**
The WindCline consists of 6 sections: blower, wide-angle diffuser, settling chamber (which includes flow conditioners), contraction, test section, and exit diffuser. It is an open-circuit blow-down type wind tunnel where ambient air is ingested by the blower and pushed through the rest of the circuit (see Fig. 1). This design prevents hot exhaust gases from soiling or overheating the fan. The blower is a Chicago Blower Corp. SQAD centrifugal direct drive 2 HP blower capable of achieving a volumetric flowrate of 3800 CFM at 1800 RPM blower speed. The blower output is controlled via a variable frequency drive to achieve a range of wind speeds through the test section (0.6 – 3 m/s in this study). The blower is additionally equipped with adjustable louvers at the inlet, which enable further adjustment of the wind speed for a given blower setting. Downstream from the blower, the air is expanded in a home-built wide-angle fiberglass diffuser before entering the settling chamber. The wide-angle diffuser has an area expansion ratio of 6.8 over the 175 cm overall length. The entrance area is matched to the blower output at 40 cm x 37 cm, and the exit area is 100 cm x 100 cm. The settling chamber entrance area is matched to the wide-angle diffuser exit area (100 cm x 100 cm) and the overall length is 22.3 cm. Within the settling chamber the flow is straightened through a 10 cm extruded aluminum honeycomb (0.64 cm cell diameter) before passing through two stainless steel M60 screens mounted 5.1 cm apart. The screens provide backpressure to the flow and break down turbulent structures larger than the screen mesh size to reduce turbulent fluctuations within the flow. Following the settling chamber, the flow is accelerated through of the square contraction section. The dimensions of the contraction are based on the fifth order polynomial [35] and the length is 100 cm with an exit area of 35 cm x 35 cm. The contraction was constructed of fiberglass and built from a wood and shaped foam mold constructed in-house.

The test section is 100 cm in length, with a 35 cm x 35 cm cross section (matched to the contraction exit). It is built from an extruded aluminum frame for easy reconfiguration. The 'standard' side panels are made from borosilicate glass for optical access and imaging (visible and near-infrared wavelengths) at elevated temperatures. The tunnel baseplate used in this study is a removable stainless-steel plate coated with a flame-proof enamel (to control the emissivity of the surface for thermal imaging - not discussed in this study). This baseplate is configured with an opening near the upstream end of the test section to accommodate a slot burner for stationary flame studies. For solid fuel experiments, the baseplate can be replaced with a sheet of combustible material or a different steel plate without the burner. The exit diffuser is constructed of welded 0.32 cm aluminum plates and has an area expansion ratio of 2.9 (entrance 35 cm x 35 cm; exit 60 cm x 60 cm) with an overall length of 145 cm. The diffuser reduces the flow speed at the wind tunnel exit to minimize pressure losses as the flow exits into the room or is passed through a HEPA filter. The choice of aluminum minimizes weight and enables easy modification to accommodate different measurement techniques downstream of the test section (e.g., emissions sampling). The diffuser exit is fitted for a HEPA filter to remove particulates during solid-fuel burning experiments. For the characterizations and preliminary tests presented here, the diffuser and HEPA filter were not installed.

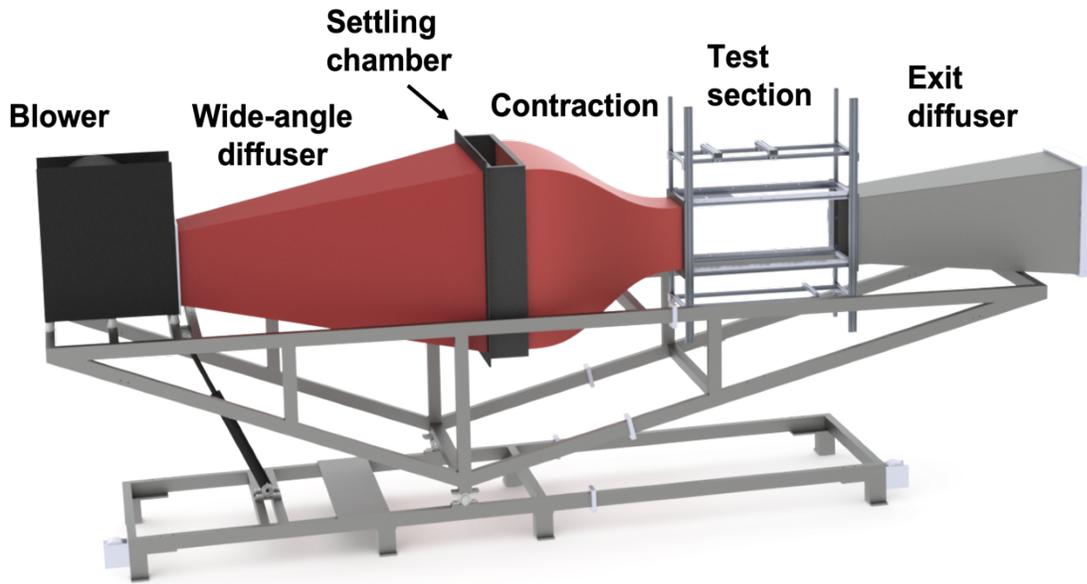

*Figure 1: CAD rendering of the WindCline with the different sections of the facility labeled.*

The wind tunnel frame was designed around a central pivot point, about which the tunnel can rotate. A hydraulic piston-cylinder assembly (Maxim 3000 psi welded hydraulic cylinder) placed below the blower is used to adjust the angle of the entire facility. The range of available testing angles is based on the extension length of the hydraulic cylinder. In the current configuration the WindCline can operate from +16° (upwards tilt, Fig. 2a) to -13° (downwards tilt, Fig. 2b). A digital protractor mounted on the baseplate of the test section or a horizontal frame beam is used to quantify tilt angle to within 0.05°. The WindCline angle is additionally stabilized via a winch system (placed on the tunnel frame near the hydraulic piston), which imposes a counter force on the piston when tightened once the desired tilting angle is reached. The winch system also acts as a fine-tuning mechanism for small adjustments to the WindCline angle.

One of the primary goals in the design of the WindCline is to enable the use of multiple diagnostic tools within both the test section and exit diffuser. Planned diagnostic tools include visual and infrared imaging systems, laser-based temperature and gas-phase species measurements (in both the near- and mid-infrared wavelength regions), velocity measurements via particle image velocimetry (PIV), hot-wire anemometry, and standard Pitot-static probe methods, and gas sampling methods to measure higher molecular weight gas-phase organic species and particulate matter (particle count, particle size distribution). The diagnostics are enabled through removable side and top panels on the test section and a diffuser that can be easily modified with various access ports for sampling. The test section side paneling will eventually include options that can incorporate specialized optical windows for working with mid-infrared sensors. Figure 3 depicts several example configurations of diagnostics that are used in an initial set of experiments presented here – laser absorption spectroscopy (LAS), visible imaging, and extractive sampling. Briefly, LAS is used to determine water vapor concentrations and absorption-weighted temperature through different portions of a diffusion flame; visible imaging is used to characterize flame structure under a variety of experimental conditions; and extractive sampling is used to measure select combustion products from a solid fuel burn experiment. Note that the configurations depicted in Fig. 3 are representative for the diagnostics and not necessarily the configurations used in this work.

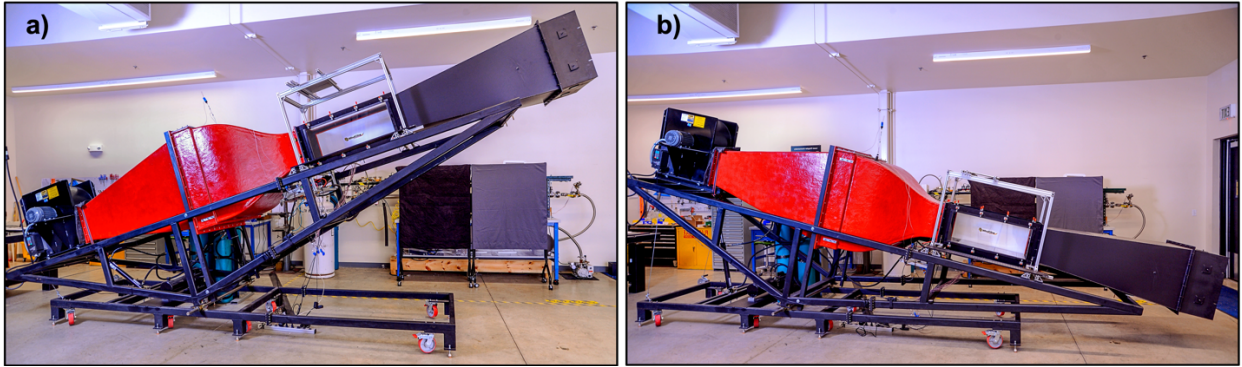

*Figure 2: WindCline at full upward (panel a) and downward (panel b) tilt angles (+16° and -13°, respectively).*

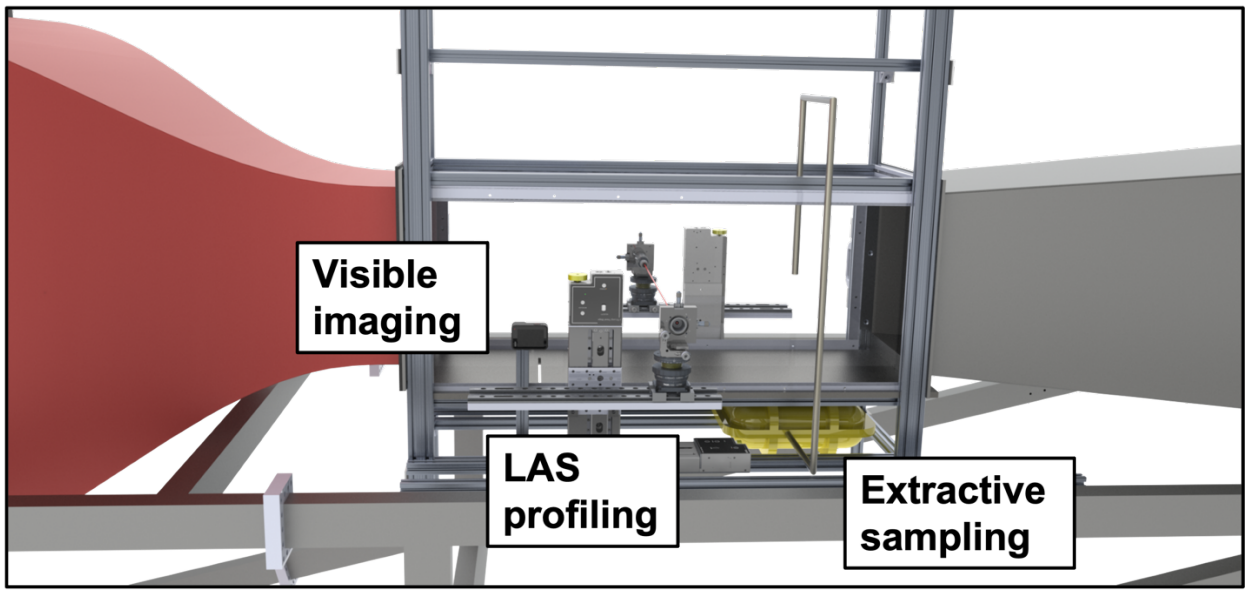

*Figure 3: Schematic showing different diagnostic configurations that can be utilized in the WindCline which are similar to the actual configurations used for the initial measurements presented in this work. Diagnostics include: laser absorption spectroscopy (LAS) measurements; visible imaging of flame structures; and extractive gas-phase sampling.*

**WindCline characterization:**
**Particle Image Velocimetry**
To characterize the incoming boundary conditions for the test section, we performed planar PIV within the test section at a position x = 15 cm upstream of the slot burner, centered in the spanwise direction. The PIV system is composed of one 2560 by 2160 pixel 16-bit dynamic range sCMOS camera from LaVision equipped with a 20-cm focal length macro lens, providing a field of view of 3.37 cm by 2.84 cm with a magnification of 13.2. A dual-pulsed 200 mJ Nd:YAG Quantel Evergreen laser (532 nm wavelength) fitted with a focusing element and a -2 cm focal length cylindrical lens is used to illuminate a vertical plane aligned in the streamwise direction. During measurements, the flow is seeded with an oil-based smoke (~0.2 μm particle size) from a ViCount Compact 1300, 1.1kW smoke machine. A programmable timing unit (PTU) from LaVision is used to synchronize the laser and camera in a double-pulsed mode allowing for the capture of two images with time separations of Dt = 220, 110, 70, 50, and 27 ms respectively for *U* = 0.55, 1.20, 1.88, 2.57, and 3.24 m/s. The two-component, two-dimensional velocity field is computed from the particle images using the DaVis 8.4 software from LaVision. We implemented a multi-pass method with the first pass using a 64-by-64-pixel interrogation window size followed by two passes at 32 by 32 pixels all incorporating a 50% overlap in interrogation windows. Note that in the final pass a Gaussian weighting was applied to the cross correlation of the interrogation windows, which were performed on the computer GPU. The resulting vector field is 161 by 136 vectors at a resolution of 47.8 vectors/cm. For each of the cases discussed here 500 velocity field snapshots were collected at 15 Hz.

**PIV Results and Discussion:**
The PIV measurements reveal that the mean freestream velocity of the WindCline facility varies linearly with the blower drive speed in the range of 1.2 m/s $< \overline{U_\infty} <$ 3.24 m/s as demonstrated in Fig. 4. The lowest speed tested, $\overline{U_\infty}$ = 0.55 m/s at a blower drive speed of 120 RPM, deviates slightly from this linear trend and shows an increase in freestream unsteadiness (i.e., turbulence intensity) and a slight change in the boundary layer profile. Error bars on the freestream velocity are calculated as the summation in quadrature of the 1-sigma standard deviation of the velocity measurements in time and the uncertainty in the PIV computed using the correlation statistics method built into LaVisions Davis 8.4 [36]. The two-component velocity unsteadiness was computed from the PIV measurements and normalized by the mean freestream velocity - this is commonly defined as the turbulence intensity. It was found to be 3.8% at the freestream speed of $\overline{U_\infty}$ = 0.55 m/s and 1.8% for each of the higher speeds in the linear response range. Time-resolved measurements using hotwire anemometry find similar values of the normalized unsteadiness (i.e., turbulence intensity) in the single-component, streamwise velocity on the centerline of the WindCline (not shown here for brevity).

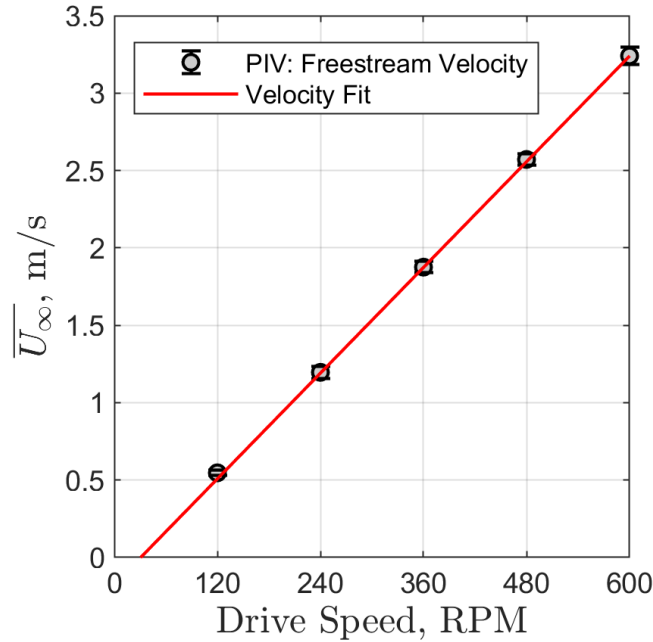

*Figure 4: WindCline mean freestream velocity with motor drive speed (RPM) (black markers) and the associated linear regression (red trace).*

The viscous boundary layer profile was also measured on the lower wall of the WindCline facility at a position x = -15 cm upstream of the slot burner. Figure 5 presents a comparison at this position of the normalized mean velocity profiles (Fig. 5a) and the two-component normalized velocity unsteadiness profiles (Fig. 5b) across the boundary layer for each of the operating speeds presented in Fig. 4. The boundary layer height, $\delta$, is estimated as the wall-normal position where the velocity recovers to 99% of the freestream value for each of the cases. This analysis demonstrates that the lowest velocity tested, $\overline{U_\infty}$ = 0.55 m/s, produces a mean velocity profile that is more representative of a laminar boundary layer under a moderate favorable pressure gradient, whereas the mean velocity profiles for the higher speeds collapse well to the theoretical Blasius solution for a zero-pressure-gradient, laminar boundary layer. Figure 5b shows that moderate velocity fluctuations, up to approximately 6% of the mean freestream speed, are present within the boundary layer for each of the freestream speeds. Further analysis of PIV data at $\overline{U_\infty}$ = 0.55 m/s reveals that instantaneous fluctuations in the boundary layer thickness, $\delta$, are present with a dominant frequency at approximately 1 Hz. It should be noted that this does not imply the boundary layer is turbulent, since each of the instantaneous PIV snapshots exhibited laminar flow within the boundary layer, but rather oscillations exist within laminar boundary layer akin to an instability wave. Additional details on the boundary layer unsteadiness are included in the Supplemental Information; however, further investigations with a time-resolved measurement system would be required to thoroughly quantify and understand this boundary layer unsteadiness. Figure 6 presents the variation of the mean boundary layer thickness, $\bar{\delta}$, mean displacement thickness, $\overline{\delta^*}$, and mean momentum thickness, $\bar{\theta}$, for each of the tested freestream velocities. At the primary freestream velocity of interest in the current study, $\overline{U_\infty}$ = 1.2 m/s (drive speed of 240 RPM), the mean boundary layer thickness is $\bar{\delta}$ = 0.93 cm thick and displays a decrease in height as the velocity is increased, which is consistent with the literature.

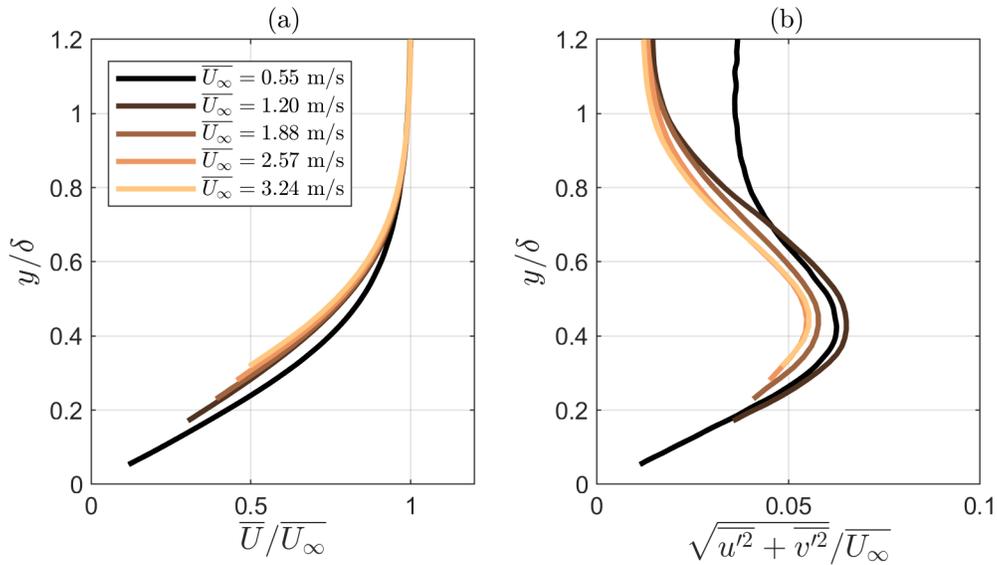

*Figure 5: Profiles of normalized mean velocity (a) and normalized two-component velocity fluctuations (b) across the boundary layer for the WindCline Facility at a position x = -15 cm upstream of the slot burner.*

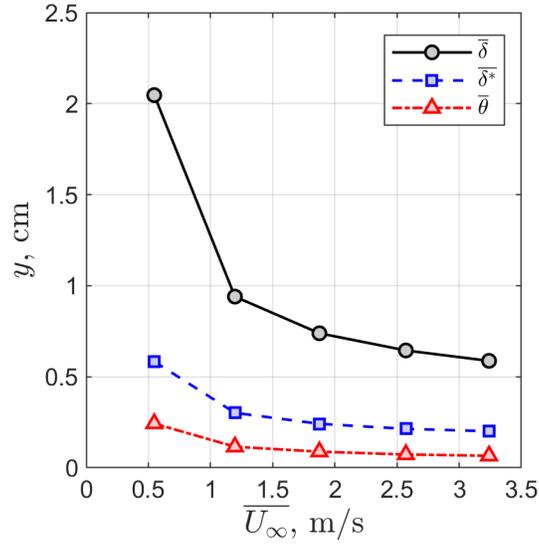

*Figure 6: Comparison of the mean boundary layer thickness ($\bar{\delta}$), mean displacement thickness ($\bar{\delta^*}$), and mean momentum thickness ($\bar{\theta}$) with freestream velocity in the WindCline facility at a position x = -15 cm upstream of the slot burner.*

Figure 7 presents a comparison of the mean velocity profile for the boundary layer at a freestream speed of $\overline{U_\infty}$ = 1.2 m/s (drive speed of 240 RPM) with the theoretical Blasius solution for a zero-pressure-gradient, laminar boundary layer during a gas-phase combustion experiment with methane flowing through the slot burner. In addition to the mean points, uncertainty bars are included, capturing both the measurement uncertainty and influence of the temporal oscillations in the boundary layer on its mean profile, as defined above for the mean centerline velocity. The measured boundary layer shows very good agreement with the theoretical solution, demonstrating that the incoming boundary layer upstream of the slot burner in the WindCline facility is indeed laminar with an effective zero-pressure-gradient. Negligible differences were observed in the upstream boundary layer properties for different inclination angles of the WindCline facility and with different operating conditions for the downstream slot burner (i.e., methane mass flow rates), thus they are not shown for brevity.

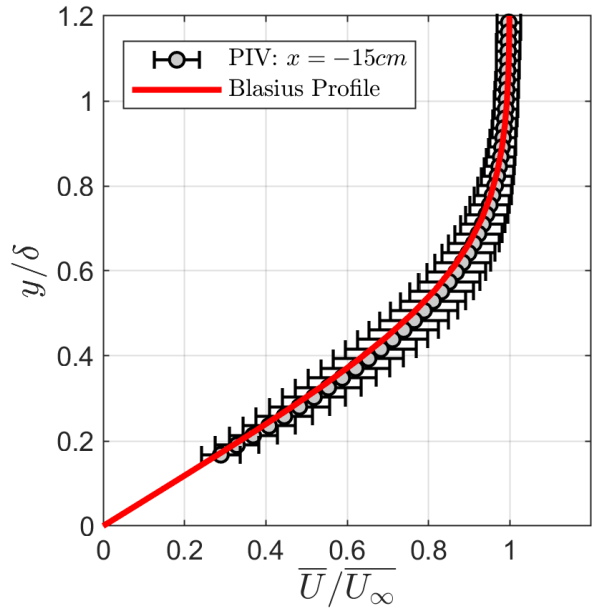

*Figure 7: Normalized mean velocity boundary layer profile with error bars capturing both the measurement uncertainty and flow unsteadiness, compared with a theoretical Blasius profile for the WindCline Facility operating at $\overline{U_\infty}$ = 1.2 m/s at a position x = -15 cm upstream of the slot burner which is combusting methane.*

**Diagnostic Suite and Initial Results:**
This section presents an overview and initial results from a series of representative diagnostic measurements completed in the WindCline. These preliminary studies are intended to demonstrate the capability of the WindCline for supporting multiple measurement platforms (diagnostics) with minimal reconfiguration of the test system. While the present experiments were all completed independently, in principle the WindCline can support simultaneous measurements, depending on the footprint and configuration required by the different techniques.

The primary goal of the WindCline facility is to create a well-controlled and characterized test system for measuring critical parameters to develop and test models of wildfire dynamics. The example experiments presented here were chosen to target flame structure (optical imaging), thermal and fluid dynamic properties (optical imaging, char visualization, laser absorption spectroscopy of gas temperature), and chemical information (laser absorption spectroscopy of species concentrations and extractive gas sampling).

**Flame structure with optical imaging:**
We conducted an initial set of experiments within the WindCline focused on visual imaging of flame structures to assess differences in bulk flame properties under varying experimental conditions. We used two Sony DSC-RX10 III Cyber-shot digital single-lens reflex (DSLR) cameras to capture video at 60 frames/s of non-premixed methane flames from the slot burner for several combinations of wind speeds (0.6, 1.2, 1.9 m/s), tilt angles (-10°, 0° ,10°), and fuel flow rates (3.0, 6.0, and 9.0 SLPM of methane; equaling a velocity of ~0.9, ~1.8, and ~2.7 m/s, respectively). We collected video from both the top-down and side view perspectives (only results from the side view camera are discussed here). We attached both cameras to rigid support structures mounted directly to the WindCline frame to ensure that the camera views relative to the test section remained consistent through different tilt angles.

Prior to testing, we mapped the spatial-length-to-pixel ratio in the flame region using a standard image calibration plate. Additionally, the inside face of the test section window on the opposite side of the camera location was painted matte black to minimize reflections of the flame in the video. We collected all data for the visual imaging tests sequentially and allowed the system to reach steady state at the desired conditions for ~10 s before starting the video recording. Each raw video recording lasted ~35 s and was trimmed to exactly 25 s during post-processing.

The side-view videos were post-processed to derive 'average' flame structure metrics over a 25 s run duration. We employed a flame image processing method similar to that presented in [37,38]. Video image processing included (1) 'averaging' the RGB pixels for each frame to return a corresponding intensity image (in grayscale); (2) determining the presence of the flame in a given pixel based on a predetermined intensity threshold (intensity > threshold = flame present; intensity < threshold = flame not present); (3) applying the threshold as a mask for each frame to return a binary image (flame present = 1; flame not present = 0); (4) averaging the video frames to return a single image with each pixel having a value between 0 and 1, which represents the fractional number of frames in which the flame was present in that pixel. We determined the threshold for (2) by visually inspecting grayscale intensity profiles from randomly selected flame images and comparing against the corresponding color frame image. A visual representation of this process is included in Fig. S4 of the Supplemental Information.

Using the average image, we created a contoured image to represent the edge of the flame, which is set to a 50% fractional threshold for the pixel values. Choosing the contour at the 50% flame/no-flame level reflects the boundary beyond which (moving away from the flame origin) the pixels contained the flame <50% of the time. We then used the contour image to derive average flame structure metrics for the different test scenarios.

In this study we determined the following properties for the different experimental scenarios: flame angle, flame horizontal extent, and flame vertical extent. We derived flame angle by calculating the midpoint between the top and bottom boundaries of the contour described in the previous section for each horizontal location downstream of the burner. This procedure generates a trend line that represents the midpoint between the top and bottom boundaries for the contour. We then fit a linear trend through this midpoint line (forcing the y-offset through the origin or burner location) and extracted the slope from this fit as the flame angle. Horizontal extent was identified as the furthest horizontal position along the contour and the flame vertical extent was identified as the maximum vertical position. Figure 8 shows the contour images generated through this analysis for three of the different test scenarios. In these plots, the different flame/no-flame percentage levels are color-coded, and the black contour outline represents the boundary of the 50% level, which is used to determine flame properties. Additionally, the results from the midpoint analysis (grey line) and the fitted trend line (black dashed line) are overlaid on each of the images.

Figure 9 gives an overview of the flame structure results, where panels a-c show the determined properties under constant fuel flow rate conditions and panels d-f are experiments with constant wind speed. We note that the small hydraulic diameter of the slot burner and the sub-unity crossflow-to-burner velocity ratio leads to a large burner-based Froude (Fr) number, corresponding to momentum-dominated jet flames in crossflow. Flow conditions and flame properties for each of the experimental configurations are provided in Table S1 of the Supplemental Information. In general, the flame angle decreases as the WindCline tilt angle increases (panels a and d), e.g., a 1.3° and 3.1° decrease in flame angle for the 1.9 m/s and 1.2 m/s wind speed cases shown in panel a, respectively. The horizontal extent increases with increasing fuel flow rate and increasing wind speed, e.g., 25 cm for angle = 0° and wind speed = 1.2 m/s (panel e) and 0.50 cm for angle = -10° and fuel flow rate = 6 SLPM (panel c). There are a few exceptions to the general trends. For example, in the lowest wind speed cases with a fuel flow rate of 6 SLPM, the flame angle remains relatively constant for the different tilt angles (panel a). Also, the horizontal extent of the flame for wind speed of 1.2 m/s, fuel flow rate of 6 SLPM, and tilt angle of 10° does not decrease as expected but increases to a level similar to that of the -10° case for the same flow conditions (panel c).

Visible imaging diagnostics can potentially be integrated as a 'standard' measurement to provide real-time/constant information on flame behavior and structure during the course of future experiments. Such a record would provide valuable insight into the flame dynamics during experimental runs, which would help with the interpretation of the results generated by other diagnostics.

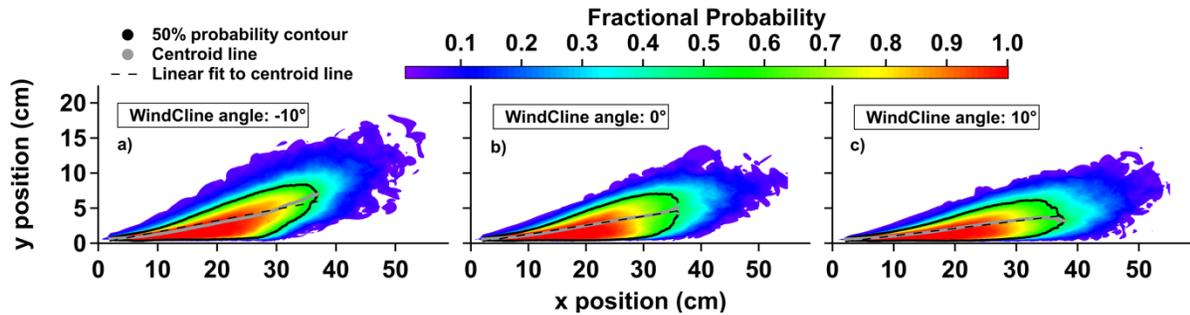

*Figure 8: Examples of the fully processed 'average' images of the flames from three test conditions – 1.2 m/s wind speed, 6 SLPM fuel flow rate, and WindCline angles of -10°, 0°, and 10° (panels a-c, respectively). The color scale represents the fractional proportion of frames in which the flame is present in each pixel, the black trace is the 50% contour bound, the grey line is the calculated mid-point between the top and bottom edges of the 50% contour for each horizontal position, and the dashed black line is the linear fit to the mid-point (grey) trace.*

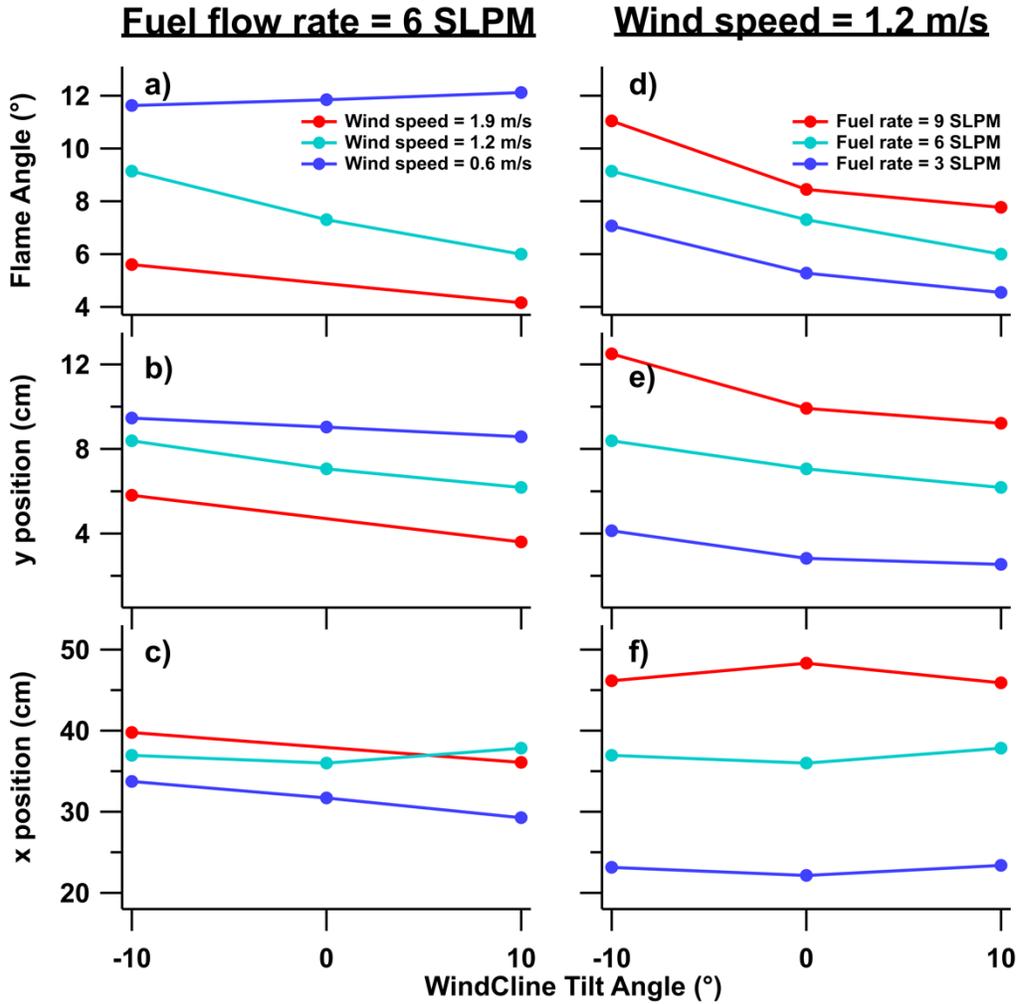

*Figure 9: Results from all visible imaging experiments. The left column (panels a-c) contains experiments with the same fuel flow rate (6 SLPM) and varying wind speed; while the right column (panels d-f) is for experiments with the same wind speed through the test section (1.2 m/s) and varying fuel flow rate. The red traces in each column represent the higher value for the variable parameters being displayed (wind speed for the left column and fuel flow rate for the right column); the aqua traces are for the 'base' conditions of fuel flow rate of 6 SLPM and wind speed of 1.2 m/s (the data for these cases are the same between the left and right columns); and the dark blue traces are for the lower value of the variable parameter. For the conditions defined by a fuel flow rate of 6 SLPM, tilt angle of 0°, wind speed of 1.9 m/s (panel b), the flame was primarily detached from the burner and not visible during video collection, and thus no analysis was completed.*

**Gas phase temperature with frequency comb laser absorption**

In another set of experiments, we implemented a laser-based diagnostic to determine gas-phase temperature and water mole fraction profiles downstream of the slot burner. We employed dual-frequency comb spectroscopy (DCS) [39,40] to measure the absorption spectrum of water vapor present in the beam path. The absorption spectrum shape exhibits unique changes with temperature and concentration, so that measurements of the spectrum at different locations can be used to determine the path-averaged water mole fraction and gas-phase temperature in the test section (and through the flame) [41]. Note that the gas-phase temperature measurements from the DCS are absorption weighted - meaning areas with higher concentrations of water vapor, such as the flame and exhaust gases, are more heavily weighted in the path-averaged temperature measurement. Since diffusion flames are fundamentally non-uniform, the reported temperatures here are an absorption-weighted average gas-phase temperature measured through the flaming areas as well as the non-combusting regions on either side of the flame.

The DCS technique uses two frequency combs (passively mode-locked femtosecond fiber lasers) [42–44] with repetition rates near 200 MHz. The short, highly stable pulse of the frequency comb results in a laser spectrum that covers a large optical bandwidth with a spectral resolution set by the repetition rate of the lasers (200 MHz or ~2 pm here), referred to as 'comb teeth'. DCS achieves high spectral measurement resolution by detecting the heterodyne interference signal created when the two frequency combs are stabilized with a slight difference in their repetition rates (~626 Hz in this study) and their output light is mixed on a photodetector. This creates unique frequency offsets between comb teeth of the two frequency combs across the entire optical spectrum. The broad bandwidth and high resolution enable high-fidelity spectroscopic measurements by fully characterizing many absorption features of the molecules of interest [45–48]. Here, we optimized the spectral output of the DCS to cover the water vapor absorption spectrum between 1390-1470 nm with a spectral point spacing of 1.8 pm in this region.

For these experiments, the DCS output was collimated (~1.2 mm diameter) and passed horizontally across the test section perpendicular to the air flow (i.e., z direction) to a focusing optic and InGaAs photodetector (Thorlabs, PDA10CF) on the opposite side of the test section. The laser transmit-and-receive optics were secured to motorized, high-precision vertical and horizontal translation stages, which we precisely controlled to maintain laser beam alignment while moving through a grid of measurement locations. A similar experimental configuration was used in [49]. For each measurement, the laser signal is averaged for 30 – 60 s to achieve an acceptable signal-to-noise ratio in the final data. This laser measurement approach produces temporally and spatially averaged values for water vapor mole fraction and absorption-weighted temperature along the laser beam path. In total, 93 measurement locations were probed using an adaptive measurement spacing, which placed the most emphasis near the surface of the test section and close to the flame. During these measurements, the flow conditions were held constant: WindCline tilt angle of 0°, wind speed of 1.2 m/s, and methane (fuel) flow rate of 6 SLPM. Figure 10(a) shows the measurement grid overlaid on the averaged image of the flame for these conditions (derived from the imaging measurements described in the previous section). Figure 10(b) shows the temperature profiles as a function of height above the baseplate for different horizontal locations. Results for the path-averaged water vapor mole fraction measurements are found in the Supplemental Information. From these measurements, we found that the path-averaged, absorption-weighted maximum temperature increases from ~685 K at the burner (0 mm) to a maximum of ~920 K at 12.8 cm (although because of the coarse horizontal measurement spacing in the downstream region the true maximum might not have been captured). The vertical location of the maximum temperature increases from $y \approx 0.3$ cm (at $x = 0$ cm) to $y \approx 2.0$ cm (at $x = 25.6$ cm), which corresponds to a linear slope of 0.066 (angle = 3.80°). This angle is approximately half of the flame angle derived from the visible image analysis presented in the previous section, highlighting the disconnect between visible emission intensity (which is produced by incandescence from soot) and gas-phase temperature within the flame. At the last horizontal location, we observed a slight decrease in the temperature between the test section baseplate and the maximum as the flame starts to separate more from the baseplate. The lowest vertical measurement (0.6 mm) for each profile remains relatively constant between 500 K - 600 K except for the measurement locations over the burner surface at $x = 0.4$ cm and $x = 0.8$ cm downstream, which are relatively cool at ~305 K and 485 K, respectively. The $x = 0.4$ cm downstream measurement location is coincident with a weld that runs along the length of the burner and could explain the noticeably lower temperature.

The DCS used in this initial experiment only spans near-infrared wavelengths that primarily cover strong water vapor absorption features. Future measurements will include the mid-infrared wavelength region (still using DCS) to substantially expand the molecular targets beyond $H_2O$ to combustion products (e.g., carbon monoxide, carbon dioxide, methane, ethane, formaldehyde, methanol, etc.) [50].

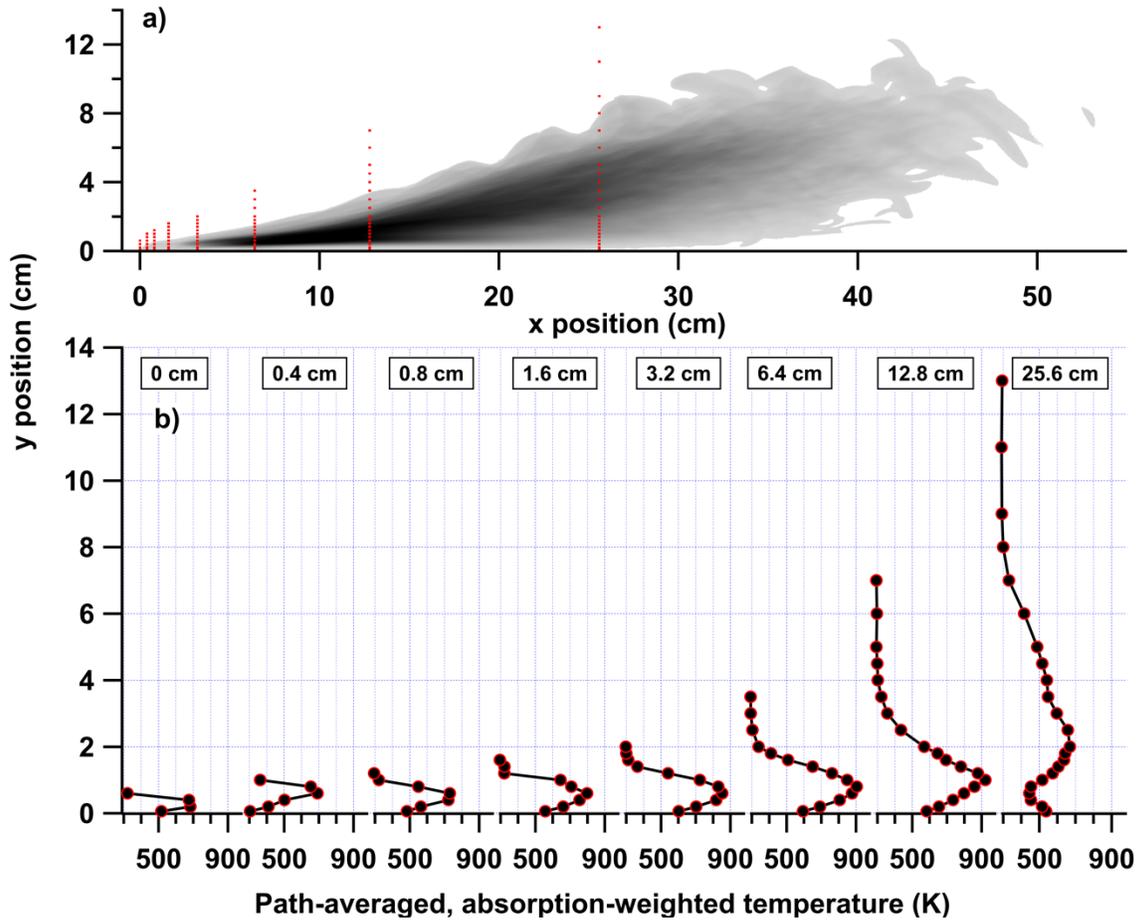

*Figure 10: Results from gas phase DCS measurements of temperature. Panel a) shows the measurement locations (red points) overlaid on an example average image of the flame. Panel b) contains the path-averaged, absorption-weighted temperature profiles for each horizontal measurement location.*

**Surface char patterns with optical imaging**

An additional optical imaging experiment was carried out to assess the charring of a solid fuel as a function of time, which can be used as an additional benchmark for computational fluid dynamics simulations. For this test, the test section baseplate was replaced with a 2.5 cm x 30.6 cm x 91.5 cm piece of planed and dried Douglas fir (oven dried at 100 °C for 2 hours). This configuration allowed the methane flame exiting the burner to interact with the sample surface to produce a char pattern. The time-dependent charring of the Douglas Fir baseplate was tracked with the same two DSLR cameras used for the prior flame imaging experiment. Both cameras were placed above the WindCline test section such that their viewing angles were effectively perpendicular to the base plate. A 470 nm (blue) optical bandpass filter was used on one of the cameras to improve the visibility of the surface char in the background of the methane flame, similar to the work of [51], and to track its progression in time at a crossflow speed of $\overline{U_\infty} = 1.2$ m/s and a methane mass flow rate of 6 SLPM.

The blue-light filtered video was post-processed to isolate the surface char from the other visible features in the field of view to quantify the evolution of the char pattern and to better understand the progression of charring in the wake of the slot burner. Specifically, the color video frames were converted to grayscale and a frame just prior to burner ignition was subtracted from each subsequent frame in the video after ignition to remove the stationary background features in the video. To minimize random noise, the differenced images were averaged over a 1 s period and a binary filter with a cutoff of -10 intensity counts was applied to each average. This processing isolated the char pattern while minimizing the error induced by the time-varying flame above the surface. Figure 11 presents these results for the full 120 s period where the surface locations are colored by the time at which surface charring was initially detected post ignition of the methane slot burner.

Surface charring was first detected at $\Delta t = 5$ s after ignition near the downstream corners of the burner ($x = 1$ cm, $z = \pm 6$ cm). The charring then evolved gradually downstream from the corners until approximately $\Delta t = 24$ s, at which point streaks of surface char began to appear within the middle of the frame between downstream positions of 5 cm $< x <$ 10 cm. Note that these cross-stream streaks aligned with the wood grain features in the specimen that was tested (see Supplemental Information Fig. S6). Over time, these streaks grew in width and length forming the final wedge shape of surface char observed in Fig. 11. No charring of the wood surface immediately downstream of the burner at a position of $x = 1$ cm was detected for the full duration of the experiments. This lack of charring suggests that the methane flame is lifted from the surface immediately downstream of the burner but reattaches at further distances. Furthermore, the most severe charring was observed along the edges of the final wedge shape, highlighting the importance of further investigating the three-dimensional structure associated with this flame-crossflow interaction. Future experiments examining char patterning and formation will focus on the effects of changing the controllable WindCline conditions (tilt angle, wind speed) as well as any effects from other wood species used for the 'baseplate' material and/or additionally including structured solid fuels within the test section.

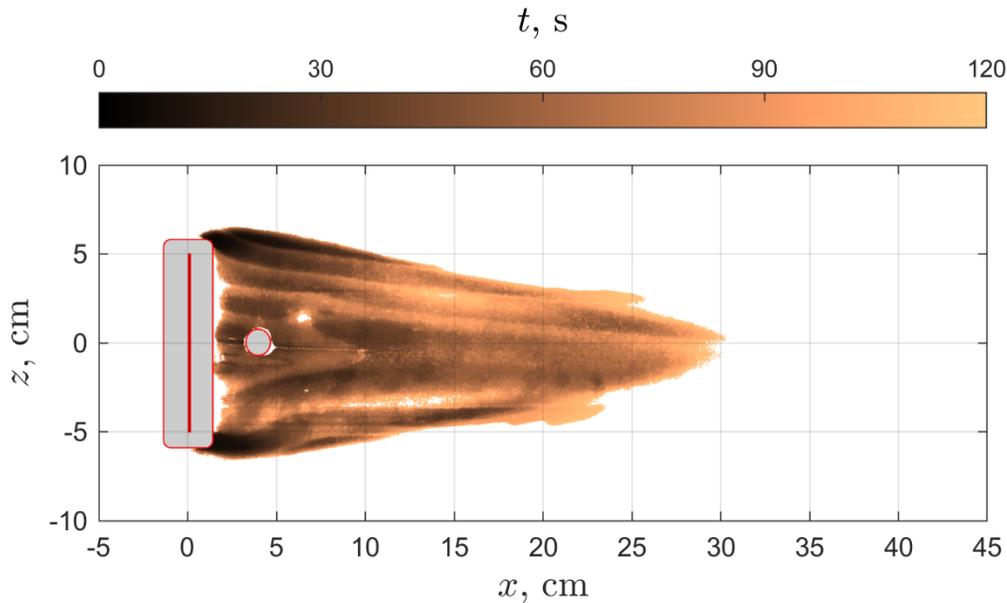

*Figure 11: Results from the char pattern experiments represented as a color scale indicating the time at which charring initiated at each spatial location on the baseplate. The small circle directly downstream of the burner is a port through which the burner was ignited; it was plugged immediately following ignition.*

**Gas composition through extractive sampling**

During the surface charring experiments, we additionally conducted gas-phase sampling of the WindCline exhaust emissions downstream of the experiment. For the sampling, we attached a t-shaped metal tube probe to the end of the WindCline test section, which collected samples from the experiment. This exhaust was extracted through the sampling tube to an emissions pod via a small pump (sampling rate set at 2 L/min). The emissions pod [52] consists of low-cost air emissions sensors including a laser scattering particulate matter (PM) sensor, an electrochemical carbon monoxide (CO) detector, a non-dispersive infrared carbon dioxide ($CO_2$) sensor, heavy and light volatile organic compound (VOC) metal oxide sensors, and temperature, humidity, and ozone sensors. The pod also includes a filter holder with a quartz fiber filter that collects PM2.5 from the air after it passes through the PM sensor. Filters can be post-processed further to determine organic and elemental carbon content and organic chemical speciation via gas chromatography and mass spectrometry, but the filters from these tests were not analyzed. Emissions sampling was performed for the duration of the charring experiment and was continued after the methane burner was extinguished. The time series of $CO_2$, CO, and $PM_{2.5}$ in Fig. 12 show that there is a transition in the primary mode of combustion prior to and after the burner is shutoff. Prior to shutoff, there is increasing $CO_2$ production associated with flaming combustion and there is only moderate CO production. After shutoff (at 12:03 pm), $CO_2$ production continues to increase and is accompanied by substantial CO production, indicative of smoldering combustion. Both products begin steadily decreasing once smoldering is finished and are unaffected by the blower state (shutoff at ~12:10 pm). $PM_{2.5}$ trends with CO, as it is primarily produced during smoldering combustion. While the measurements from this preliminary set of experiments are somewhat limited in scope, the capabilities of extractive sampling are scalable. Not only can the analysis method be changed for the samples (i.e., use of a more extensive technique, such as gas chromatography and mass spectrometry to analyze the extracted sample), but also the sample location can be changed to target different metrics. For instance, different sampling locations can be placed along the exit diffuser to assess chemical speciation in the exhaust gas, which could provide information on chemical reactions occurring as the exhaust gas expands and cools in the diffuser. Depending on the overall diagnostic suite and goals for any given measurement effort, the primary consideration for the extractive sampling methods is ensuring that acquiring the sample does not impact the flame dynamics (if that is a parameter of interest).

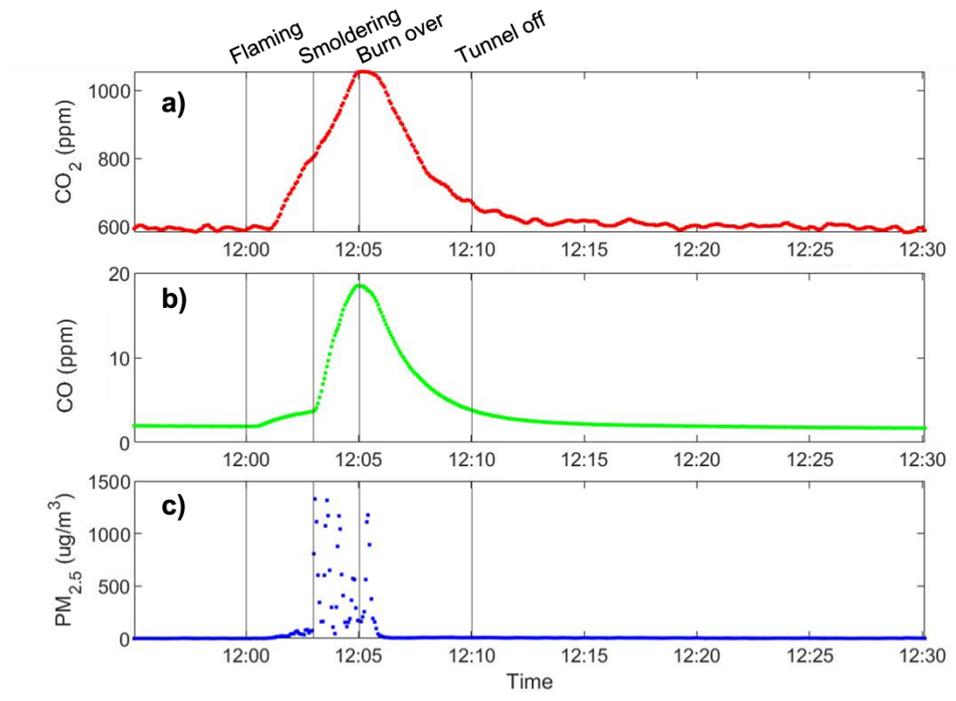

*Figure 12: Time series of the gas sampling results during char experiments including $CO_2$ (panel a), CO (panel b), and $PM_{2.5}$ (panel c). The vertical lines indicate the status/state of the burn during the experiment – burner lit at 12:00 pm, burner shutoff at 12:03 pm, flame is completely extinguished at 12:05 pm, and the crossflow is stopped at 12:10 pm.*

**Conclusion:**
We present the development and characterization of a testbed for small-scale fire dynamics, the 'WindCline'. The design specifications and fabrication of the WindCline were specifically chosen to enable control over many critical parameters important to understanding wildfire processes (e.g., wind speed, terrain angle, fuel loading) and boundary conditions to enable the careful study and modeling of these systems. Additionally, the WindCline is designed to support a broad suite of advanced diagnostics for probing conditions within the test section as well as the capability to be reconfigured for additional diagnostics. Similarly, the diffuser construction supports the addition of diagnostics to probe the air downstream of the test section. The characterization of the WindCline included a detailed study using particle image velocimetry (PIV) to understand the velocity fields present within the WindCline test section. This analysis demonstrated consistent results over the range of tested conditions and good agreement between measured and theoretical values for a laminar boundary layer profile on the floor of the test section where the flame interaction with the cross flow is focused.

Initial experiments included visible imaging and characterization of flame structure at different run conditions, measurement of gas phase thermodynamic properties (temperature and water vapor mole fraction) through a diffusion flame under wind-loaded conditions, and an analysis of char growth and gas-phase combustion products of a solid wood sample under wind-loaded conditions. Visible imaging experiments demonstrated flame behavior consistent with expectations for measured parameters as a function of the run conditions; e.g., increasing flame angle with decreasing WindCline tilt angle, increasing horizontal flame extent with increasing fuel flow rate. However, some trends did not hold at the lowest wind speed tested (0.6 m/s), which would indicate a change in the factors driving these parameters. The discrepancies at the lower wind speeds could also be related to the decreased stability (relative to higher wind speeds tested) of the facility in this regime as determined by the PIV measurements. The results from the gas phase temperature measurements and the charring experiments will be key to aiding the development of computational models. These experiments provide information on important parameters of combustion in the WindCline test section; the measured temperature and water vapor distributions can be used to tune models to ensure that chemistry, fluid dynamic, and heat transfer physics of the process are being correctly modeled. The charring experiments can also be re-created through modeling to verify that solid fuel interactions and combustion processes are adequately derived. Additionally, relationships between parameters such as visible image intensity and temperature can be derived and then used as additional metrics for validating modeling results.

The control over the boundary conditions within the WindCline and the ability to characterize how the boundary conditions impact flame behavior lay the groundwork for developing computational models of fire properties that can be transferred to more complex conditions.

## Supplemental Information
### Boundary Layer Unsteadiness

Moderate unsteadiness of the laminar boundary layer was observed from the analysis of the planar PIV measurements for all conditions and cases tested. More specifically 500 snapshots were taken at sample frequency of 15 Hz producing a sample period of $T = 33.33$ s with a temporal resolution of $\Delta t = 0.067$ s for each case. Figure S1 presents velocity vector fields of the laminar boundary layer at relative time instants of t = 2.73 s (a), and t = 4.93 s (b) from the start of data collection where the background contour presents the local instantaneous magnitude of the flow speed, and the white line delineates the instantaneous edge of the boundary layer at each time instant. Note that this specific case is at a mean freestream speed of $\overline{U_\infty} = 1.2$ m/s while the slot burner is combusting methane at a flow rate of 6 SLPM. The two instants in time presented in Fig. S1 were chosen to highlight the minimum (a) and maximum boundary layer thicknesses encountered during the initial 10 s of data collection where Fig. S2(a) presents a time trace of the boundary layer profile and Fig. S2(b) presents the histogram for all 500 snapshots both at a streamwise position of $x = -15$ cm. The time trace clearly shows that the boundary layer unsteadiness is not fully time-resolved with the current measurements, however the histogram demonstrates that the collected statistics are still sufficient to analyze the mean properties of the boundary layer. Note that a fast Fourier transform of the full-time trace of the boundary layer thickness (not shown here for brevity) revealed a dominate frequency around 1 Hz. Finally, Fig. S3 presents the overlay of all 500 instantaneous normalized boundary layer profiles at the same streamwise position, $x = -15$ cm, for comparison with the theoretical Blasius profile for a steady, zero-pressure gradient boundary layer. There is clear variation in the boundary layer profile in time, however the collection of profiles in combination aligns well with the Blasius solution. Furthermore, the spread in the profiles is also accurately captured by the error bars presented in Fig. 7 for the mean boundary layer profile.

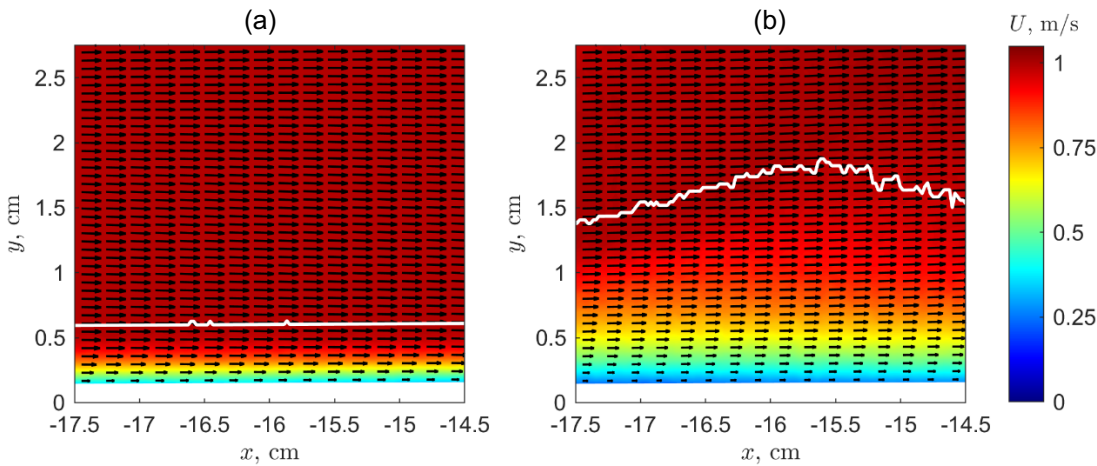

*Figure S1. Instantaneous velocity vector fields overlayed on contours of flow speed at relative time instants of t = 2.73 s (a), and t = 4.93 s (b). The white line delineating the instantaneous boundary layer edge, δ, where $U = 0.99\ U_\infty$.*

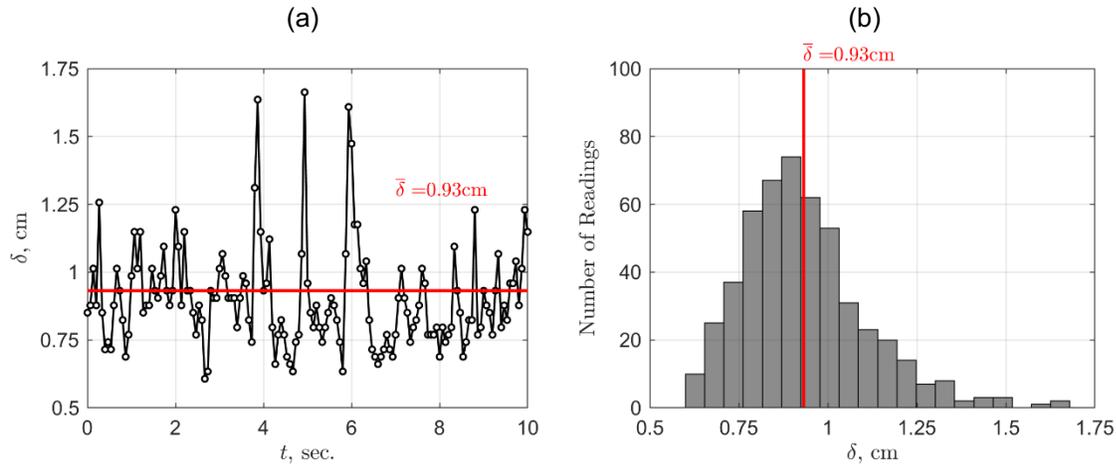

*Figure S2. Time variation for the first 10 s of data collection, approximately one third of the full sample period (a) and histogram of full sample period (b) of the boundary layer thickness, δ, at a streamwise position x = -15 cm upstream of the slot burner.*

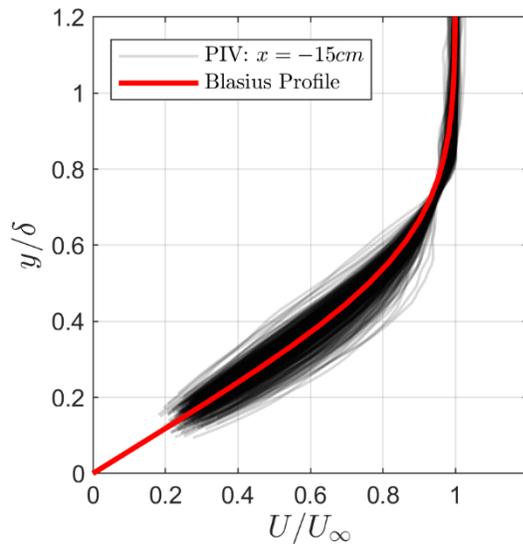

*Figure S3. Instantaneous boundary layer profiles for all 500 PIV snapshots at a position x = -15 cm upstream of the slot burner compared with the theoretical Blasius boundary layer profile.*

**Flame structure with optical imaging**

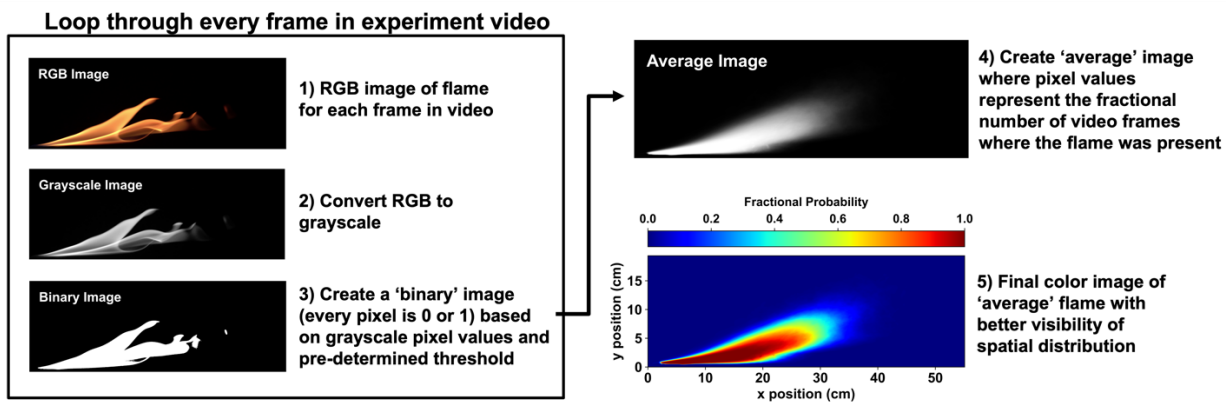

*Figure S4. Pictorial representation of image processing steps.*

*Table S1. I Information and results for the different optical imaging experimental conditions.*

| Case No. | Fuel Flow Rate (SLPM) | WindCline tilt angle (deg) | Wind speed (m/s) | Slope | Angle (°) | Horizontal Extent (cm) | Vertical Extent (cm) |
|---|---|---|---|---|---|---|---|
| 1 | 6 | -10 | 0.5 | 0.206 | 11.6 | 33.7 | 9.6 |
| 2 | 6 | 0 | 0.5 | 0.210 | 11.8 | 31.7 | 9.0 |
| 3 | 6 | 10 | 0.5 | 0.215 | 12.1 | 29.3 | 8.6 |
| 4 | 6 | -10 | 1.2 | 0.161 | 9.1 | 37.0 | 8.4 |
| 5 | 6 | 0 | 1.2 | 0.128 | 7.3 | 36.0 | 7.1 |
| 6 | 6 | 10 | 1.2 | 0.105 | 6.0 | 37.8 | 6.2 |
| 7 | 6 | -10 | 1.9 | 0.098 | 5.6 | 39.8 | 5.8 |
| 9 | 6 | 10 | 1.9 | 0.073 | 4.2 | 36.1 | 3.6 |
| 10 | 3 | -10 | 1.2 | 0.124 | 7.1 | 23.1 | 4.1 |
| 11 | 3 | 0 | 1.2 | 0.092 | 5.3 | 22.1 | 2.8 |
| 12 | 3 | 10 | 1.2 | 0.080 | 4.5 | 23.4 | 2.5 |
| 13 | 9 | -10 | 1.2 | 0.195 | 11.0 | 46.1 | 12.5 |
| 14 | 9 | 0 | 1.2 | 0.149 | 8.4 | 48.3 | 9.9 |
| 15 | 9 | 10 | 1.2 | 0.136 | 7.8 | 45.9 | 9.2 |

**Gas phase $H_2O$ measurements**

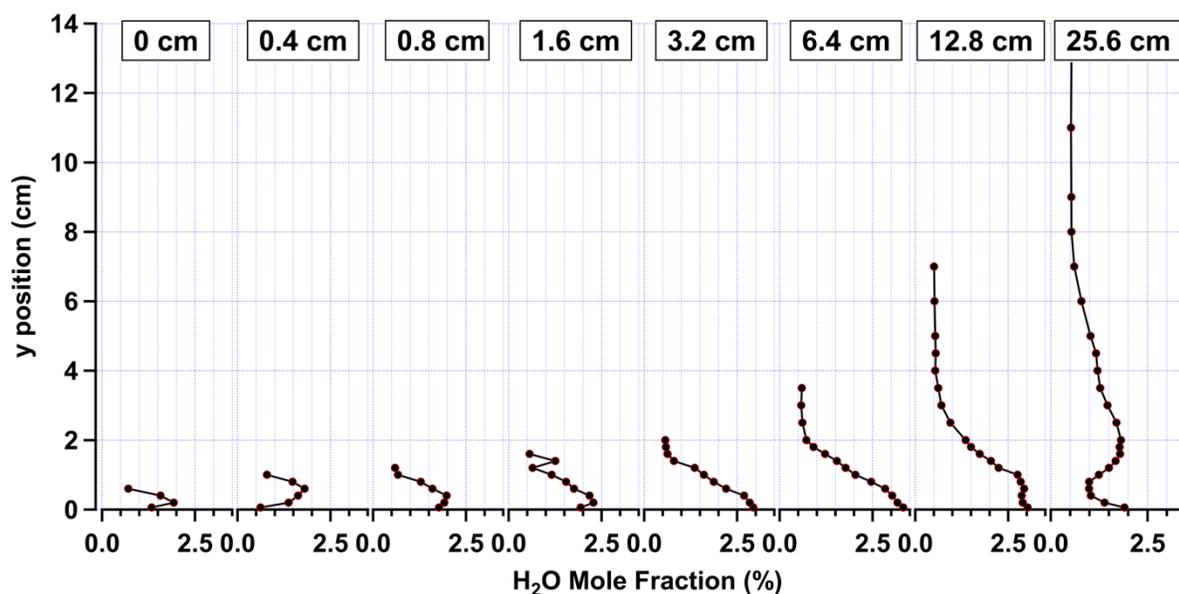

*Figure S5. Vertical profiles of the path-averaged water mole fraction measured by the DCS at each of the horizontal locations downstream of the slot-burner. These measurements are produced from the same data used to derive the path-integrated, absorption-weighted temperature measurements presented in the main body of text.*

**Char experiments**

Figure S6 presents the spatially calibrated color images demonstrating the instantaneous flame and surface char at a time $t = 60.3$ s post ignition of the methane slot burner. Figure S6(a) presents the color image from the camera with no optical filtering, while Fig. S6(b) displays the nearly identical frame from the camera with a 470 nm bandpass (blue light) filter installed on its lens. These images demonstrate that the blue light filter moderately reduces the optical interference of the flame luminosity but is not able to eliminate the brightest regions, which varied both spatially and temporally during the experiment.

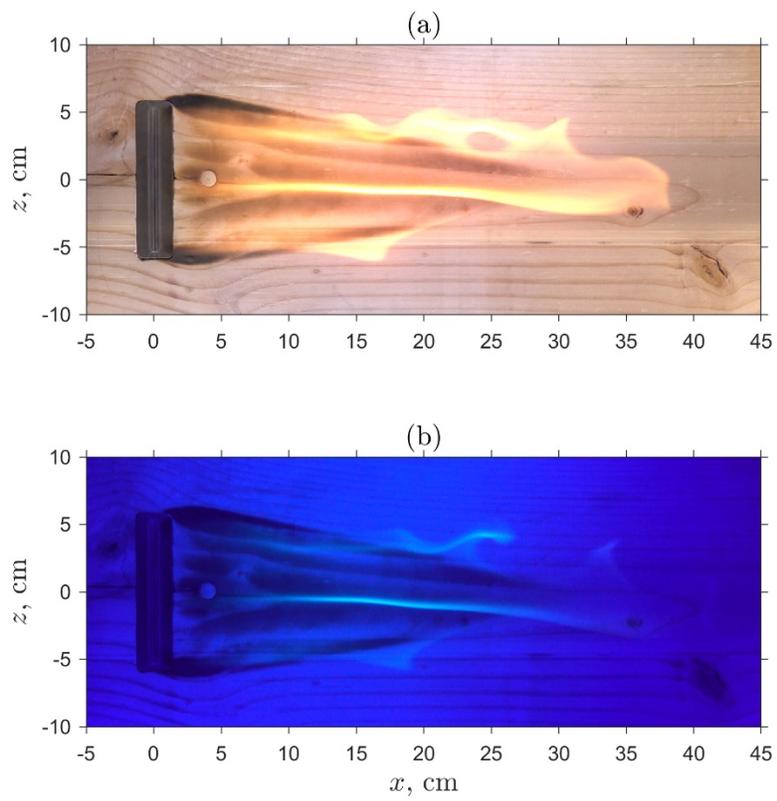

*Figure S6: Visualization of the experimental surface charring of a Douglas Fir baseplate in the WindCline facility at t = 60.3 s after ignition of the methane burner.*

**Funding**
This work was supported by the Strategic Environmental Research and Development Program (W912HQ-16-C-0026 and W912HQ-20-C-0065).

**Conflicts of Interest**
The authors have no conflicts of interest to disclose.

**Author Contributions**
**Amanda S. Makowiecki**: conceptualization, methodology, project administration, resources, supervision, writing/original draft preparation, writing/review & editing
**Sean C. Coburn**: conceptualization, data curation, formal analysis, investigation, methodology, project administration, resources, supervision, visualization, writing/original draft preparation, writing/review & editing
**Samantha Sheppard**: conceptualization, data curation, formal analysis, investigation, methodology, resources, writing/review & editing
**Alexandra Jaros**: methodology, resources
**Robert Giannella**: methodology, resources
**Christopher Kling**: methodology, resources
**Brendan Bitterlin**: methodology, resources
**Abdul Dawlatzai**: investigation, methodology, resources
**Timothy Breda**: investigation, methodology, resources
**Eric Kolb**: investigation, methodology, resources, writing/original draft preparation
**Caelan Lapointe**: conceptualization, methodology
**Sam Simons-Wellin**: conceptualization, methodology, writing/review & editing
**Hope A. Michelsen**: conceptualization, methodology, writing/review & editing
**John W. Daily**: conceptualization, methodology, writing/review & editing
**Michael Hannigan**: conceptualization, methodology, supervision, writing/review & editing
**Peter E. Hamlington**: conceptualization, funding acquisition, methodology, project administration, supervision, writing/review & editing
**John Farnsworth**: conceptualization, formal analysis, funding acquisition, investigation, methodology, project administration, resources, supervision, visualization, writing/original draft preparation, writing/review & editing
**Gregory B. Rieker**: conceptualization, funding acquisition, methodology, project administration, supervision, writing/original draft preparation, writing/review & editing

**Data Availability Statement**
Data included in this study are available from the authors upon reasonable request.

**Dedication**
Dedicated to the memory of Caelan Lapointe